\documentclass[11pt]{article}

\usepackage{jcappub,soul,color}

\usepackage{graphicx,color,soul}
\usepackage{latexsym}
\usepackage{amsmath,amssymb}        

\title{Stability of BEC galactic dark matter halos}

\author{F. S. Guzm\'an,}
\author{F. D. Lora-Clavijo,}
\author{J. J. Gonz\'alez-Avil\'es,}
\author{F. J. Rivera-Paleo.}
\affiliation{	Instituto de F\'{\i}sica y Matem\'{a}ticas, \\
	Universidad
              Michoacana de San Nicol\'as de Hidalgo. \\ Edificio C-3, Cd.
              Universitaria, 58040 Morelia, Michoac\'{a}n,
              M\'{e}xico.}

\emailAdd{guzman@ifm.umich.mx}
\emailAdd{fadulora@ifm.umich.mx}
\emailAdd{javiles@ifm.umich.mx}
\emailAdd{friverap@ifm.umich.mx}

\date{\today}

\abstract{
In this paper we show that spherically symmetric BEC dark matter halos, with the $\sin r/r$ density profile, that accurately fit galactic rotation curves and represent a potential solution to the cusp-core problem are unstable. We do this by introducing back the density profiles into the fully time-dependent  Gross-Pitaevskii-Poisson system of equations. Using numerical methods to track the evolution of the system, we found that these galactic halos lose mass at an approximate rate of half of its mass in a time scale of dozens of Myr. We consider this time scale is enough as to consider these halos are unstable and unlikely to be formed. We provide some arguments to show that this behavior is general and discuss some other drawbacks of the model that restrict its viability.}


\begin{document}

\maketitle


\section{Introduction}

An interesting dark matter candidate is a spinless ultralight boson. The idea originates at cosmic scale, where a scalar field in effective theories is assumed to play the role of dark matter, and in fact what happens is that such candidate mimics cold dark matter at cosmic scale with its mass as the free parameter, which in order to fit the mass power spectrum of structures its value is fixed to an ultrasmall value $m\sim 10^{-23} - 10^{-22}$eV$/c^2$  \cite{MatosUrena2000}. This candidate is thus a potential solution to the problems of overabundance of small structures attached to the CDM model, and has shown to have all the benefits of $\Lambda$CDM model at cosmic scales \cite{cinvestav}. The idea of the scalar field in an effective theory works fine at cosmic scales, however once the structures are assumed to evolve on their own gravity, the Einstein-Klein-Gordon system reduces to the time-dependent Schr\"odinger-Poisson system of equations \cite{GuentherThesis,Choi,Wai,BernalTesis}.

A key point is that the resulting time-dependent Schr\"odinger equation is considered to represent the evolution of a Bose Condensate at zero temperature, in the mean field approximation, including a point to point self-interaction between pairs of particles \cite{ChavanisHarko}. The gravitating version of the condensate would be described by the Schr\"odinger-Poisson coupled system of equations, where the gravitational potential acts as the trap binding the condensate.
The density of probability calculated with the solution wave function is interpreted as the density of dark matter structures, in particular galaxies and therefore the source of Poisson equation. In this sense, Schr\"odinger equation plays the role of the Gross-Pitaevskii equation of Bose condensates \cite{GrossPitaevskii} and the whole system of equations ruling the gravitating condensate can be called  the Gross-Pitaevskii-Poisson system (GPP).

Aside of being a proposal to solve the overabundance of small structures, the model is appealing because the critical temperature of Bose condensation is extremely high, corresponding to scale energies of $T_c \sim $TeV for boson masses of $m\sim 10^{-22}$eV \cite{RoblesMatos2012}. In this way structures are expected to form at early times, which could explain the existence of early galaxies as a consequence of the condensation process.

However at present cosmological time, the drops resulting after the condensation should be consistent with observations, specifically, a galactic dark matter model must show: 1) dark matter halos must be late time attractor solutions, such that the resulting density profiles are the expected result of the evolution of an initial dark matter fluctuation with a rather general initial profile, 2) it must explain galactic rotation curves (RCs) and 3) preferably the resulting dark matter halos should not have the cusp-core problem. 

For instance, the standard model is associated to a late time attractor NFW density profile that explains very well RCs, however shows the core-cusp problem. Related to the BEC dark matter halos, some advances have been achieved in this directions that we describe here. The GPP system of equations reads

\begin{eqnarray}
i\hbar \frac{\partial \tilde{\Psi}}{\partial \tilde{t}} &=& -\frac{\hbar^2}{2m}\tilde{\nabla}^2 \tilde{\Psi} + \tilde{V}\tilde{\Psi} +\frac{2\pi \hbar^2 \tilde{a}}{m^2} |\tilde{\Psi}|^2\tilde{\Psi}, \nonumber\\
\tilde{\nabla}^2 \tilde{V} &=& 4 \pi G m |\tilde{\Psi}|^2, \label{eq:SPcompleta}
\end{eqnarray}

\noindent where in general $\tilde{\Psi} = \tilde{\Psi}(\tilde{t},\tilde{{\bf x}})$, $m$ is the mass of the boson, $\tilde{V}$ is the gravitational potential acting as the condensate trap and $\tilde{a}$ is the scattering length of the bosons. When the wave function is assumed to depend harmonically in time and spherical symmetry is assumed, the wave function can be written as $\tilde{\Psi} = e^{i\tilde{E} \tilde{t}/\hbar}\tilde{\psi}(\tilde{r})$, and the system of equations above reduces to a Sturm-Liouville eigenvalue problem for $\tilde{\psi}(\tilde{r})$, provided boundary conditions on the wave function. Solutions to such eigenvalue problem have been constructed numerically for instance in \cite{Ruffini}, and further studied in \cite{Gleiser1998,SinJin} and in fact already applied as dark matter halos \cite{SinJin,GuzmanUrena2004}, including arbitrary values of $\tilde{a}$ \cite{GuzmanUrena2006}. Two types of solutions were constructed, ground state solutions characterized by a wave function without nodes, and excited state solutions characterized by wave functions with nodes. It has been shown numerically that excited state solutions decay into ground state solutions \cite{Arbey,GuzmanUrena2006}. Therefore, among these types of solutions, the ones potentially usable as dark matter halos would be the ground state solutions.

These ground state solutions have been shown to be stable against radial  \cite{GuzmanUrena2006} and non-radial perturbations \cite{BernalGuzman2006a},  and have shown properties of late-time attractors \cite{GuzmanUrena2006}. The later property indicates that for very general initial profiles, either a solution of the eigenvalue problem or not, if the evolution of a wave function is ruled by system (\ref{eq:SPcompleta}), it will approach a ground state solution. Moreover, these configurations have been shown to be virialized for arbitrary values of $\tilde{a}$.

These ground state configurations  show interesting density profiles at the central parts of halos, however, when this happens, rotation curves are not satisfactory because the configurations are very compact. In fact, excited state solutions show better RCs that extend along galactic scales, however, as mentioned above, these configurations are unstable \cite{SinJin,GuzmanUrena2004}. An interesting attempt to rescue this model consists in assuming that galactic halos are mixed ground-excited states, which certainly improve the RC and conform stable configurations. This may eventually be the solution to the problem of RCs \cite{BernalUrena2012}.

An alternative and totally different approach to BEC dark matter halos was presented in \cite{BoehmerHarko2007} by B\"ohmer and Harko, where the authors apply the Madelung representation of the wave function and construct a hydrodynamical version of the GPP system. Under the assumption of hydrostatic equilibrium and in the Thomas-Fermi limit, the authors arrive to a Lane-Emden type of equation, which is solved assuming a barotropic equation of state with solutions for polytropic index $n=1$. The resulting density profile of their solution is

\begin{equation}
\tilde{\rho}_{BEC}(\tilde{r}) = \tilde{\rho}_{BEC}^{c} \frac{\sin(\pi \tilde{r}/\tilde{R})}{\pi \tilde{r}/\tilde{R}}, \label{eq:densityprofile}
\end{equation}

\noindent where $\tilde{R}$ is the radius at which $\tilde{\rho}_{BEC}(\tilde{R})=0$, defined as the size of the galaxy, and $\tilde{\rho}_{BEC}^{c}$ is the central density of dark matter. These two quantities are fitting parameters of RCs when compared with observations. Furthermore $\tilde{R}=\pi \sqrt{\frac{\hbar^2 \tilde{a}}{Gm^3}}$, where again $m$ is the mass of the boson and $\tilde{a}$ its scattering length \cite{ChavanisHarko}. This approach works fine for an arbitrary condensate, in particular in \cite{BoehmerHarko2007} the authors fix the value of $\tilde{a} \sim 5.77 \times 10^{6}$fm, measured in the laboratory  for $^{87}$Rb, then by fitting some galactic rotation curves the boson mass results to be $m \sim $eV. These parameters rule out this boson gas because  the mass if not ultralight and this would be inconsistent with the mass power spectrum fits \cite{MatosUrena2000}.

This is why a modified version of the model was introduced. Using this same density profile, but considering the boson mass $m\sim 10^{-23}$eV$/c^2$, different studies have been carried out.  For instance in \cite{RoblesMatos2012}, the authors use the density profile (\ref{eq:densityprofile}) to fit RCs for a sample of dwarf and LSB galaxies, and at the same time found that the model might be a solution to the cusp-core problem \cite{HarkoCC, RoblesMatos2012, RoblesMatos2}. Thus at first sight this modified version looks very promising because the core density profile also fits very well.

We want to stress that equilibrium solutions of the GPP system of equations are different from  the equilibrium solutions constructed by B\"ohmer and Harko. A clear difference is that the density profile for ground state equilibrium configuration of the GPP system decay exponentially in space and  is different from that in (\ref{eq:densityprofile}).  This of course implies different RC profiles. In comparison, the B\"ohmer-Harko profile with ultralight boson mass seems to fit much better, rotation curves and cores \cite{RoblesMatos2012, RoblesMatos2} than the ground state solutions.

We show here however, that  B\"ohmer-Harko profiles (\ref{eq:densityprofile}) with ultralight boson mass  are unstable. 

Before showing it, a word has to be said about the stability of ground state solutions of the GPP  eigenvalue problem. Such time independence  immediately implies that the density of probability is time independent and furthermore the gravitational potential is time-independent. These implications on the time independence do not guarantee that the solutions are stable, in fact excited solutions constructed under the same equilibrium assumptions are unstable \cite{GuzmanUrena2006}. Going further, in \cite{GuzmanUrena2006} it was shown that for negative $\tilde{a}$ there are stable and unstable branches, even for ground states solutions.
We thus stress that ground state configurations for arbitrary non-negative $\tilde{a}$ are stable under spherical and non-spherical perturbations \cite{GuzmanUrena2006,BernalGuzman2006a}.

On the other hand, in the B\"ohmer-Harko approach, the time-independence assumptions are applied in the Madelung frame, where profile (\ref{eq:densityprofile}) is a solution. As far as we can tell, the stability of this profile has never been shown. There are two simple options to probe its stability properties. The first one is the perturbation analysis of the solution and a second one could consist in the evolution of the solution. We choose the later and apply the same kind of techniques used to probe the stability of the ground state equilibrium solutions, as in \cite{GuzmanUrena2006,BernalGuzman2006a}.

In this way, we translate back the profile (\ref{eq:densityprofile}) into the original full time-dependent  GPP system (\ref{eq:SPcompleta}) in spherical symmetry and study its evolution.

The paper is organized as follows. In section 2 we present the methods used to evolve the BEC halos, in section 3 we show results corresponding to the stability results of (\ref{eq:densityprofile}) halos and finally in section 4 we present a discussion about the problems and status of the BEC dark matter model at galactic scales.

\section{Evolution of BEC halos}

{\it Units.} We rescale the quantities in equations (\ref{eq:SPcompleta}) such that  
$\hat{\Psi} = \frac{\sqrt{4\pi G}\hbar}{mc^2}\tilde{\Psi}$,
$\hat{r} = \frac{mc}{\hbar}\tilde{r}$,
$\hat{t} = \frac{mc^2}{\hbar}\tilde{t}$,
$\hat{V} = \frac{\tilde{V}}{mc^2}$,
$\hat{a} = \frac{c^2}{2mG}\tilde{a}$, so that the numerical constants $\hbar,~\hbar^2/m,~2\pi \hbar^2/m^2,~4\pi Gm$ do not appear in (\ref{eq:SPcompleta}). Additionally, we set our code units allowing accurate calculations, using the invariance of system (\ref{eq:SPcompleta}) under the transformation 
$t=\lambda^2 \hat{t}$, 
$r = \lambda \hat{r}$, 
$ \Psi = \hat{\Psi}/\lambda^2$,
$V = \hat{V}/ \lambda^2$,
$a = \lambda^2 \hat{a}$, 
 for an arbitrary value of the parameter $\lambda$ \cite{GuzmanUrena2004}. 
 We write the operator $\nabla^2$ in (\ref{eq:SPcompleta}) in spherical coordinates and assume all the variables depend only on $r$ and $t$. Our variables in code units for the spherically symmetric case in spherical coordinates, obey the following system of equations
 
 \begin{eqnarray}
i \frac{\partial \Psi}{\partial t} &=& -\frac{1}{2}\left( \frac{\partial^2 \Psi}{\partial r^2} + 4 \frac{\partial \Psi}{\partial r^2}\right) + V\Psi +a |\Psi|^2 \Psi, \nonumber\\
\frac{\partial^2 V}{\partial r^2} + 4 \frac{\partial V}{\partial r^2} &=& | \Psi |^2, \label{eq:SPcode}
\end{eqnarray}

\noindent where the second term in the Laplacians is a regular version of $\frac{2}{r}\frac{\partial }{\partial r} = 4\frac{\partial }{\partial r^2}$ where the later is a first derivative with respect to $r^2$; this avoids numerically the singularity at $r=0$ in spherical coordinates. We use the resulting coupled system of equations to track the evolution of a given initial wave function corresponding to a galaxy with appropriate parameters in (\ref{eq:densityprofile}) that fit actual galactic RCs.

{\it Evolution of the B\"ohmer-Harko galactic halos for ultralight boson mass.}  We solve the system (\ref{eq:SPcode}) in the domain $r\in[0,r_{max}],~ t \ge 0$ as follows. Schr\"odinger equation is an evolution equation for $\Psi$, which feeds Poisson equation, which in turn is required backwards in Schr\"odinger equation. We use a finite differences approximation of the equations on a uniform grid covering the domain. For the evolution equation we use a method of lines (MoL) integrated in time using a third order iterative Crank-Nicholson integrator, with second order spatial stencils; during the intermediate steps of the MoL, we solve Poisson equation using a sixth order Numerov algorithm \cite{Numerov} with a monopolar boundary condition for $V$. Because the galaxy is considered to be localized with a surface defined at $r=R$, we integrate the system in a domain $r \in [0,r_{max}]$ where we choose different  values of the outer boundary in order to check our results are independent of the domain $r_{max}=3R,~4R$. We implemented a sponge in the farthest part of the domain covering a region $r\in [r_{max}-R/2,r_{max}]$, which captures density of probability escaping from the gravitational potential and prevents such density from reflecting back into de numerical domain and contaminate the calculations. 

{\it Initial data.} In order to dynamically evolve the system, we start with an initial wave function consistent with (\ref{eq:densityprofile}). The Madelung transform is such that $\Psi(r,t) = \sqrt{\rho(r,t)}e^{iS(r,t)/\hbar}$, where $S$ is a phase with units of action, which under the assumption of hydrostatic equilibrium demands $\nabla S =0$ and assuming a harmonic time dependence of $\Psi$ the phase $S$ can be written as $S=-Et$, where $E$ is the eigenenergy of the system \cite{ChavanisTF}, thus at $t=0$ we have $S=0$ . The profile (\ref{eq:densityprofile}) then provides a simple recipe to construct the wave function of the dark matter halo.

The size of the galaxy is $R$, defined by the condition $\rho_{BEC}(R)=0$, the first zero of the density. 
As said before, we set an extended domain $r_{max}>R$, which allows us to monitor the motion of density of probability toward regions with $r>R$.  Therefore, the initial wave function is set to

\begin{equation}
\Psi(r,t=0) = \left\{ \begin{array}{ll}
				\sqrt{\rho_{BEC}(r)} & ~~r < R,\\
				0 &  ~~R \le r \le r_{max}
			\end{array} \right .
			\label{eq:initialdata}
\end{equation}

\noindent where $\rho_{BEC}(r)$ is given by (\ref{eq:densityprofile}).

Finally, we fix an appropriate $\lambda$ allowing accurate calculations in terms of resolution, numerical domain and amplitudes.
 
 {\it Diagnostics.} We track important physical quantities on the fly during the evolution. Among the most important ones are the expectation values of the kinetic, gravitational and self-interaction energy that decide whether or not the system is initially unbounded or no, and its possible evolution toward a final estate. Explicitly such expectation values are calculated as
 
 \begin{equation}
 K = -\frac{1}{2}\int \Psi^* \left( \frac{\partial^2 \Psi}{\partial r^2} + 4 \frac{\partial \Psi}{\partial r^2}\right) r^2 dr,~~
 W = \frac{1}{2}\int V |\Psi|^2 r^2 dr,~~
 I = \int a |\Psi|^4 r^2 dr.
 \end{equation}
 
 \noindent As an extra check, with these definitions we have probed that ground state equilibrium configurations \cite{GuzmanUrena2006} are virialized and by solving (\ref{eq:SPcode}) we made sure such solutions remain stationary.
 

\section{Results}

\begin{figure*}[htp]
\includegraphics[width=8cm]{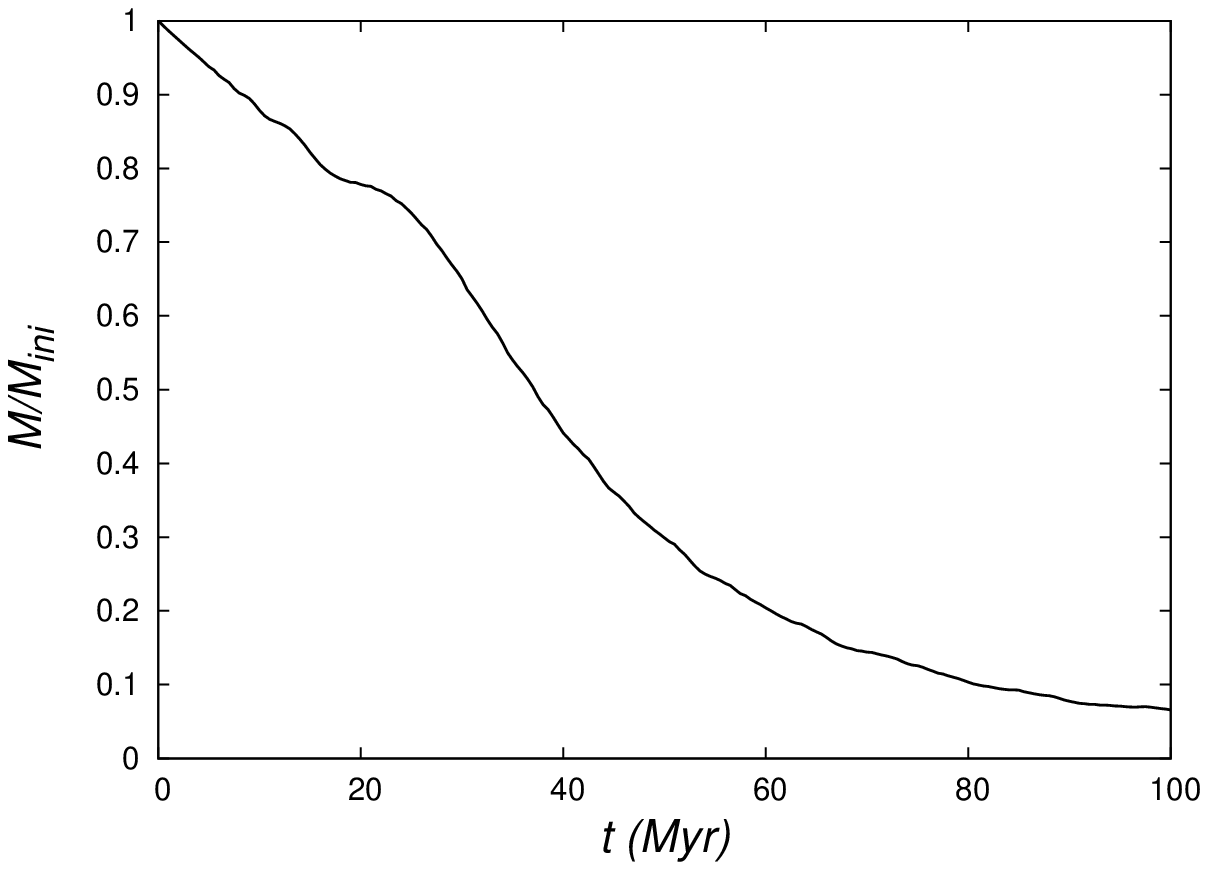}
\includegraphics[width=8cm]{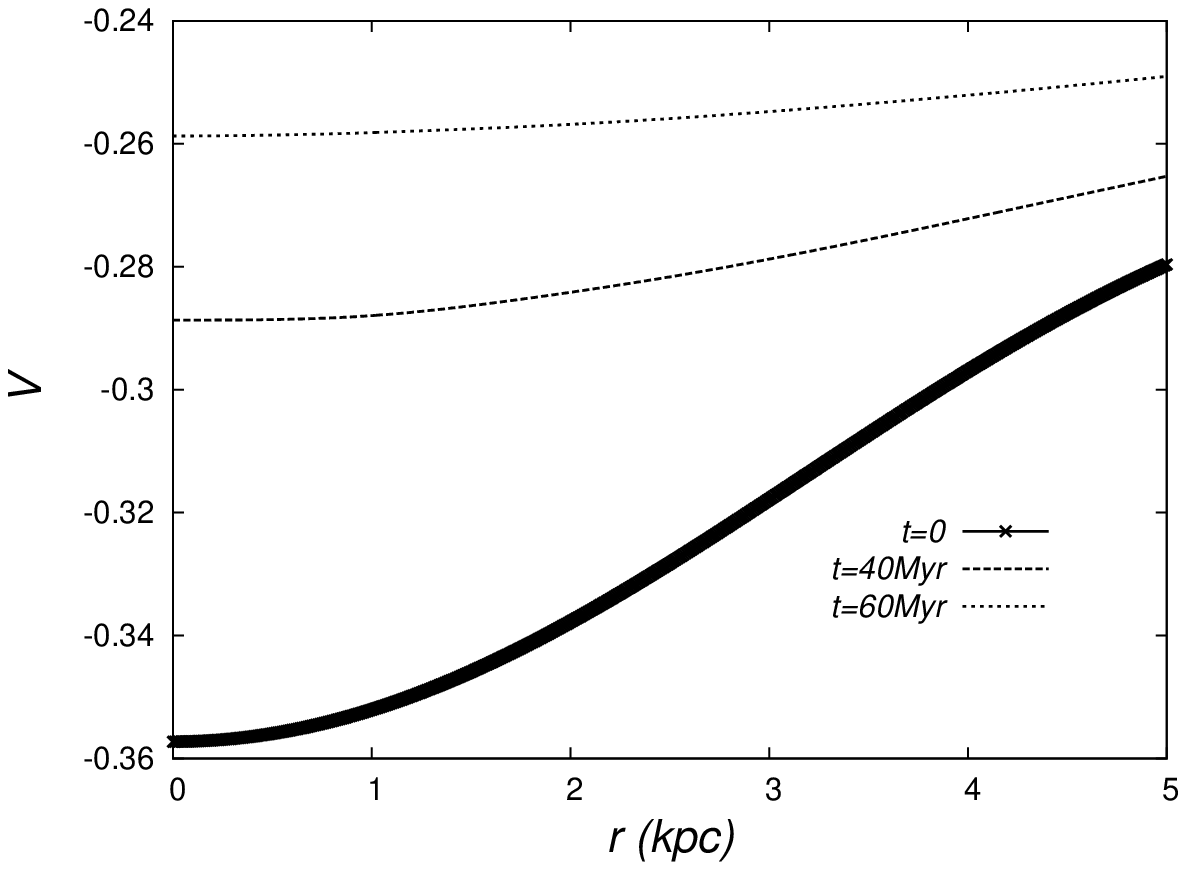}
\includegraphics[width=8cm]{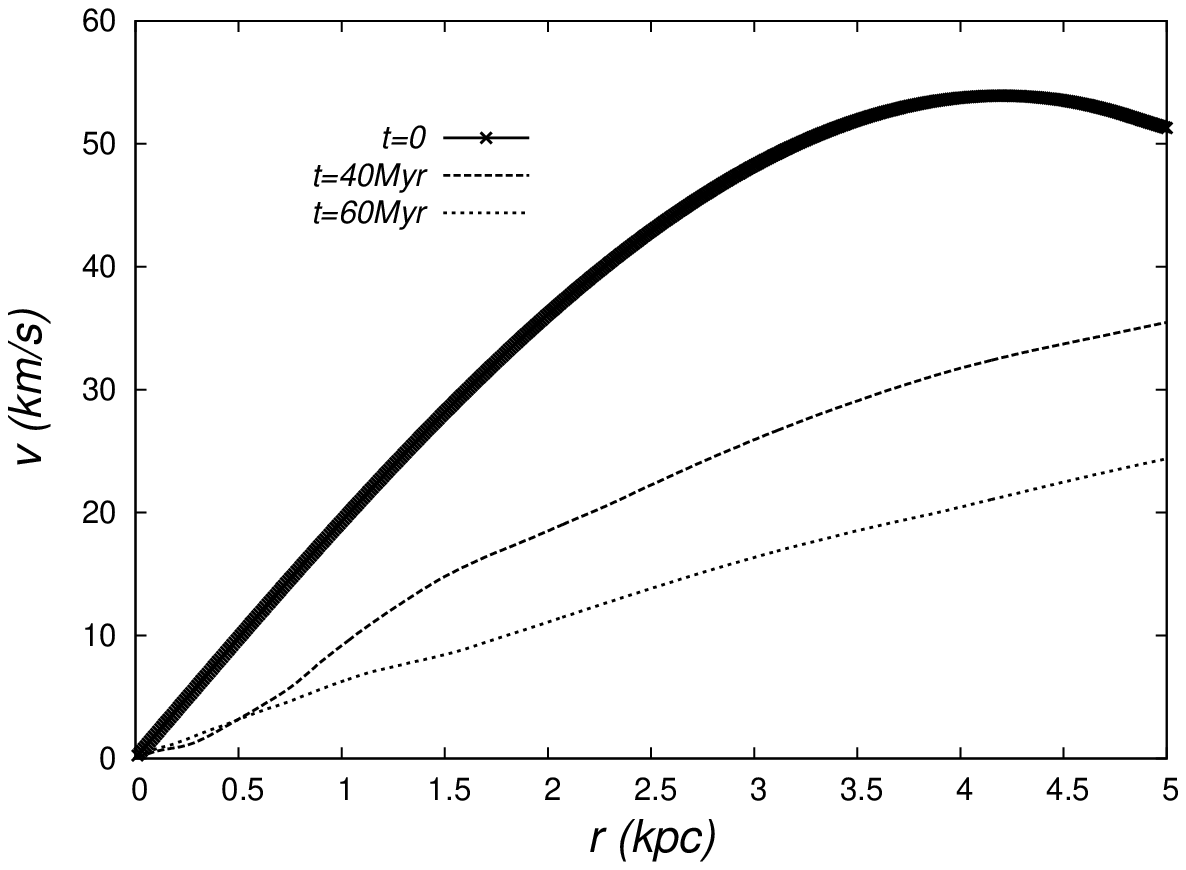}
\includegraphics[width=8cm]{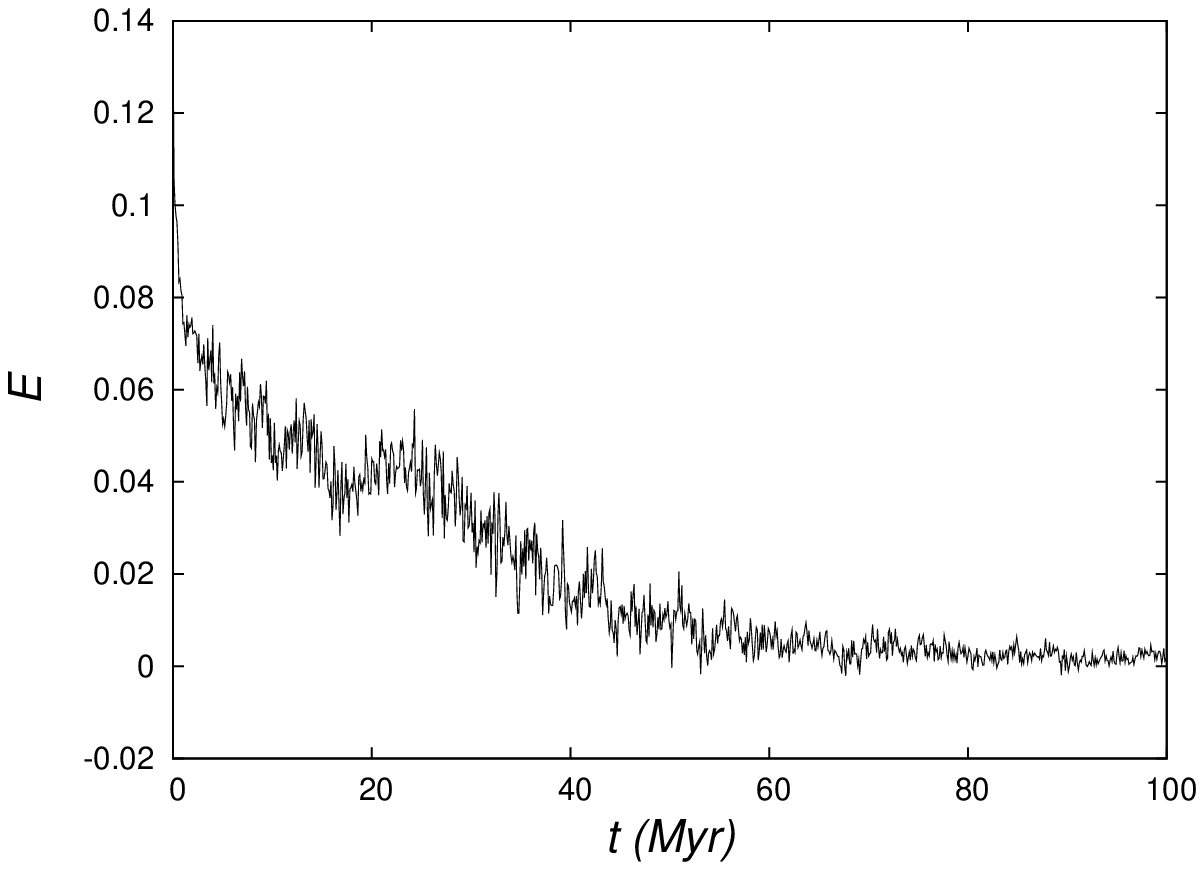}
\caption{\label{fig:case1} Galaxy ESO3050090. We show snapshots of the gravitational potential at various times in code units and show that it does not remain time independent. We also show the rotation curve and how it quickly distorts in time. We show the mass as a function of time contained within $r=R$ and show how it quickly decreases. Additionally we show in code units the total energy $E=K+W+I$ and show it is positive at the beginning and all the way during our evolutions; the fact that it is positive initially already indicates that the system is unbounded and should be dispersed away. The parameters fitting the galactic rotation curves in code units are $R=4.81$ and $\rho^{c}_{BEC}=0.001013$; also in code units the value of the scattering length is $a=1.17$.}
\end{figure*}

\begin{figure*}[htp]
\includegraphics[width=8cm]{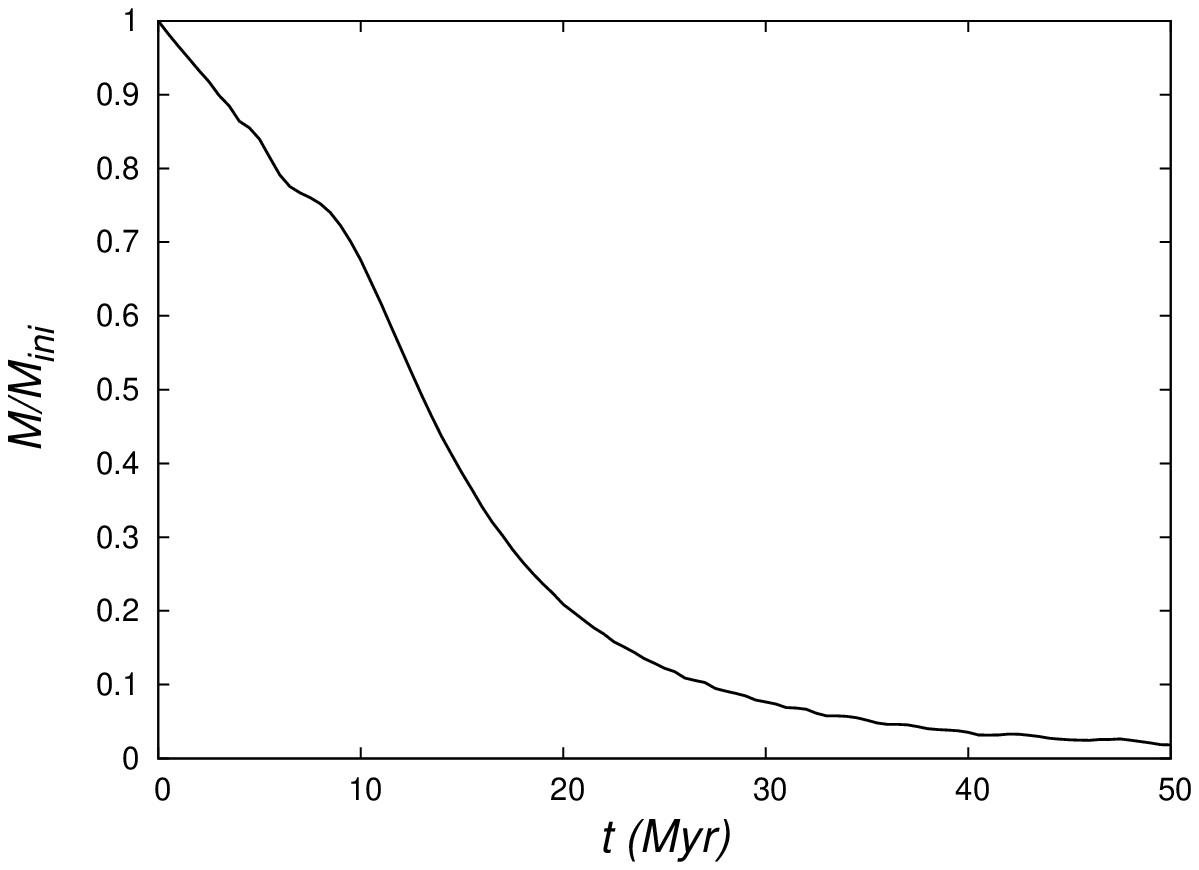}
\includegraphics[width=8cm]{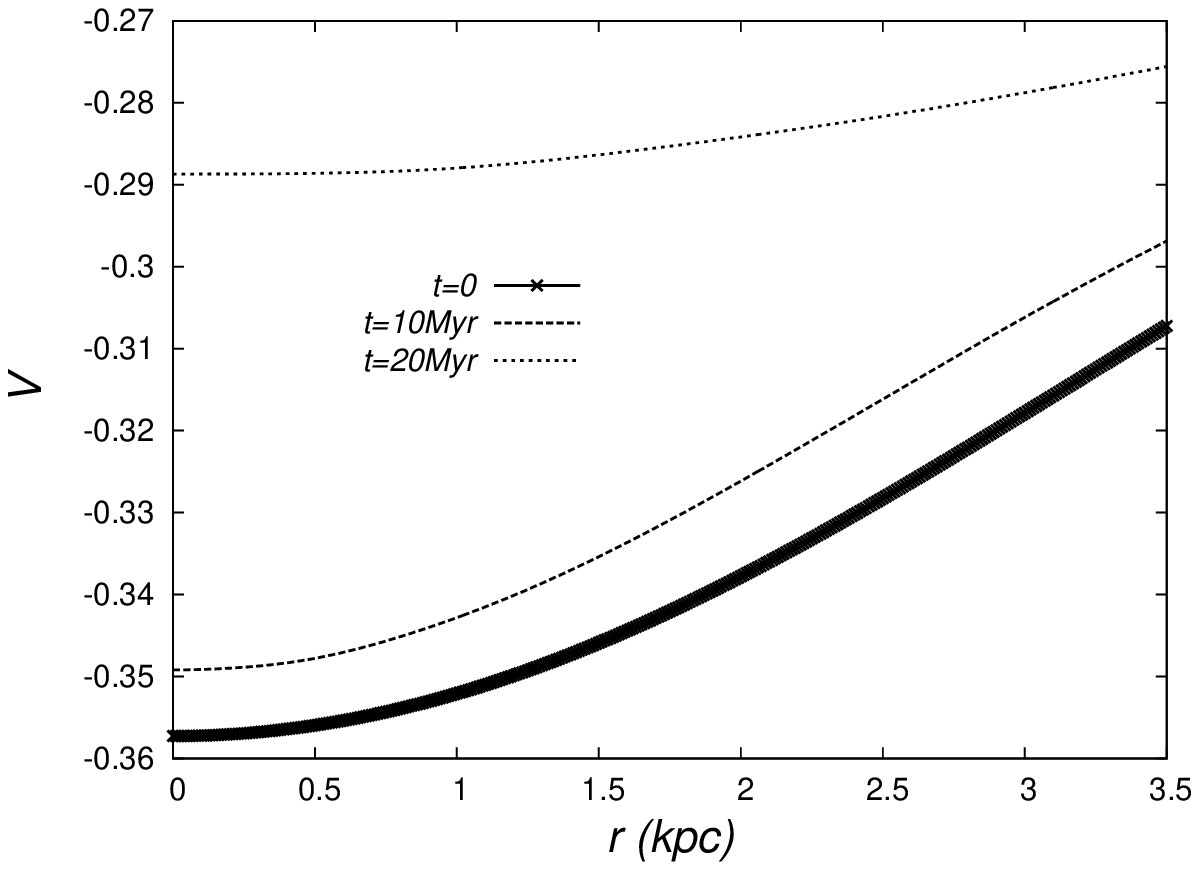}
\includegraphics[width=8cm]{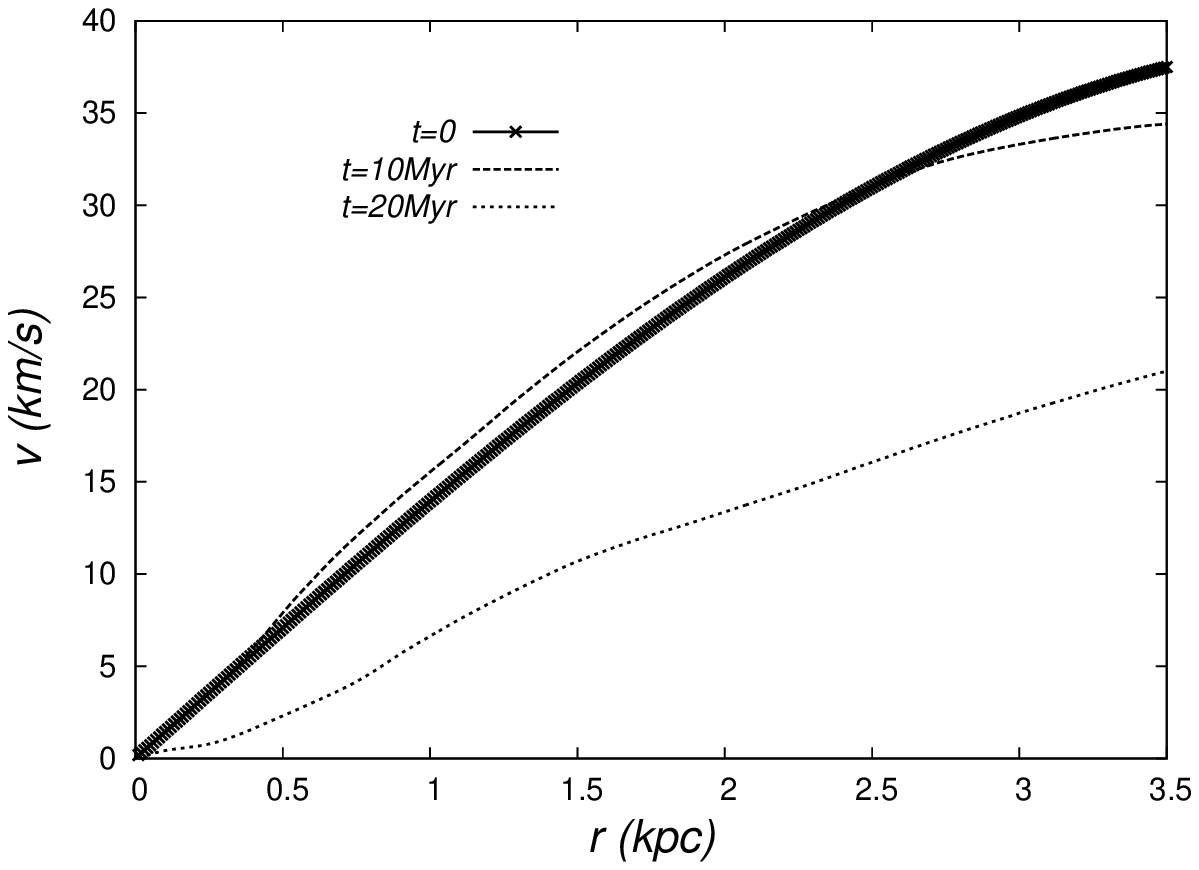}
\includegraphics[width=8cm]{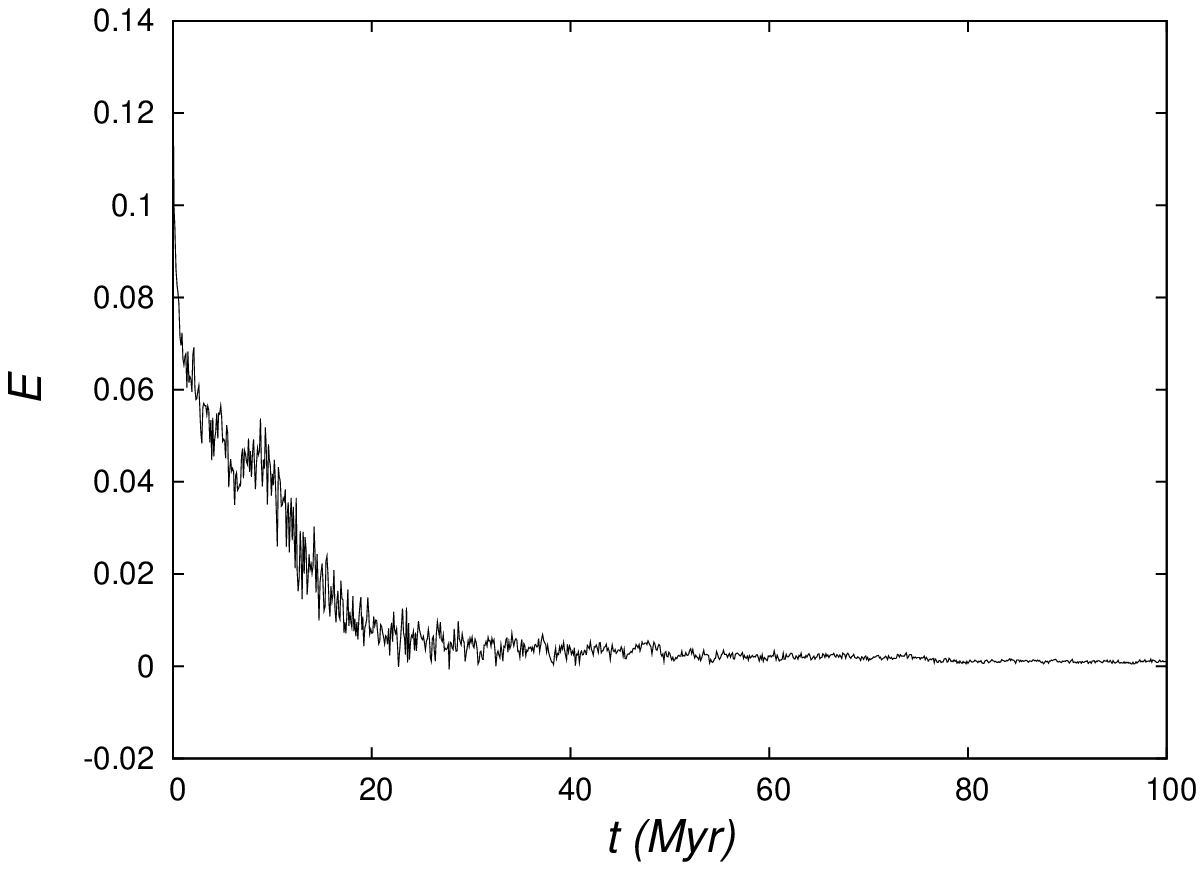}
\caption{\label{fig:case2} Galaxy ESO1870510. The behavior is totally similar to that of the first galaxy. In this case the parameters are $R=2.929$, $\rho^{c}_{BEC}=0.002328$ and $a=0.43$ in code units.}
\end{figure*}

We present results for two different profiles corresponding to RC fits of ESO3050090 in Fig. \ref{fig:case1} and galaxy ESO1870510 in Fig. \ref{fig:case2}, which show accurate RC and core fits \cite{RoblesMatos2012}. The parameters found for these two galaxies are respectively $\tilde{R}=4.81$kpc and $\tilde{R}=2.93$kpc. With these values of the galactic size and a given value of the boson mass $m$ one calculates $\tilde{a}=\frac{Gm^3 \tilde{R}^2}{\hbar^2 \pi^2}$. In Table \ref{tab:pars} we show the value of $\tilde{a}$ for these two galaxies in units of meters $[mt]$ for the mass values $m=10^{-23}$eV$/c^2$ which corresponds to a good bosonic dark matter candidate.

The hatted variables are simply calculated using the relations that define them. The self-interaction parameter $\hat{a}$ is shown also in the table for the two galaxies and the two values of the boson mass mentioned.

Finally, we fix the scale invariance parameter $\lambda$ using the spatial coordinate (it could also be the mass or the central density) such that $\lambda = \frac{\hbar}{mc}\frac{r}{\tilde{r}}= \frac{\hbar}{mc}\frac{R}{\tilde{R}}$. We want $\tilde{R}=R$ in order to use a numerical domain with coordinate values as in kpc; for that we only write down the factor $\frac{\hbar}{mc}$ in kpc. For $m=10^{-23}$eV$/c^2$ we have $\lambda = \frac{\hbar}{mc}[kpc]=0.0006399$. 
The values for $a$ are shown in the table as well.

The second physical parameter fitting the RCs is the central density. This quantity is scaled for $\hat{\rho} = |\hat{\Psi}|^2$ and $\rho = |\Psi|^2$ according to the formula

\begin{equation}
\rho = \frac{\hat{\rho}}{\lambda^4} = \frac{1}{\lambda^4} \frac{4\pi G \hbar^2}{m^2 c^4} \tilde{\rho}
\end{equation}

\noindent which for $m=10^{-23}$eV$/c^2$ and $\lambda = 0.0006399$ implies for the two galaxies the central density in code units: $\rho_{c, ESO3050090}=0.001013$ and $\rho_{c, ESO1870510}=0.002328$. With $R$ and $\rho_c$ in code units we define the initial profile for the wave function according to (\ref{eq:initialdata}). 

\begin{table}
\begin{center}
\begin{tabular}{|c|c|c|}\hline
&  ESO3050090 & ESO1870510 \\\hline
$\tilde{a}[mt]$ 	& $7.55\times 10^{-80}$ & $2.8\times 10^{-80}$ \\
$\hat{a}$ 	& $2.86 \times 10^6$ & $1.06 \times 10^6$ \\
$a$	& 1.17 & 0.43 \\\hline
\end{tabular}
\end{center}
\caption{\label{tab:pars} Based on the parameters obtained for the RC fits $R = 4.81$kpc and $R = 2.93$kpc for the two galaxies in \cite{RoblesMatos2012}, and considering an ultralight boson mass $m=10^{-23}$eV$/c^2$ we fix $\lambda=0.0006399$. We show the values of the scattering length in meters ($\tilde{a}$), hatted and code units.}
\end{table}


In order to show the stability properties of these halos we may apply an explicit perturbation, say using the addition of a Gaussian shell to (\ref{eq:initialdata}) in order to trigger an  instability. However, we found that it was not necessary, because the total energy $E=K+W+I$ is positive at initial time and thus the system unbounded, and in turn implies the configurations are expected to disperse away. The small perturbations due to the finite differences truncation errors suffice to trigger a time development of the system and we did not need to apply explicit perturbations. 

Provided the total energy is positive we expect the configuration to release the energy through the emission of particles and tend toward a relaxed state (see. e.g. \cite{SeidelSuen1990,GuzmanUrena2006}). We show this happens by monitoring the mass contained within $r=R$; this implies that the gravitational potential -which should be time-independent for a realistic galaxy- diminishes and the RCs distort. We also show the evolution of the total energy of the system and show it decreases. This shows that the configurations are unstable and tend toward an unknown configuration. 

In order to have an idea of the value of the excess of energy in these two examples, we write the total energy in physical units $\tilde{E}=(\hbar c^3 / 4\pi G m)\lambda^3E$, which for the value used for $\lambda = 0.0006399$, the galaxy ESO3050090 has the initial energy $6.91\times 10^{48}$J, whereas for the galaxy ESO1870510 the excess of energy is $6.45\times 10^{48}$J.

In order to compare with an independent estimate and check that positive total energy is expected for theses galaxies, the excess of energy can be estimated using a given ansatz for a self-gravitating configuration, for instance the Gaussian ansatz, that considers the density of probability has a Gaussian shape \cite{ChavanisTF}. For such ansatz $\tilde{K}=\frac{N\hbar^2}{2m R^2}$, $\tilde{W}=-\frac{3}{4}\frac{GM^2}{R}$ and $\tilde{I}=\frac{3}{2}\frac{N^2 \hbar^2 a}{mR^3}$, where $N$ is the number of bosons. For $m=10^{-23}$eV$/c^2$ it happens that $N_{ESO3050090}=3.43\times 10^{98}$ and $N_{ESO1870510}=1.17\times 10^{98}$. With these numbers of particles and the values of the other involved constants, the total energy in each case is $\tilde{E}_{ESO3050090}=+1.27\times 10^{48}$J and $\tilde{E}_{ESO1870510}=+5.57\times 10^{48}$J. These values are not exactly those we calculate with the exact profile (as expected), however the sign of the total energy is the relevant output of this estimate. In fact,  the positive sign of the energy will preserve for the original model by B\"ohmer-Harko for a mass $m \sim $eV. It is surprising that this calculation was not done after fitting rotation curves in both, the eV boson in \cite{BoehmerHarko2007} and the ultralight boson in \cite{RoblesMatos2012}. This would immediately indicate that this sort of configurations are unstable since the beginning.

Finally, even if unstable, it is interesting to estimate the time it takes a configuration to change appreciably its properties. Concerning the mass loss by the initial configurations, we show in Fig. \ref{fig:case1} that the mass of the halo decays to half of its initial value at about 40Myr, and in the second example in Fig. \ref{fig:case2} the time scale is  15Myr. These time scales are considerably short compared with the $10^9$yr life-time scale that galaxies seem to show. Therefore, these halos are unstable and have  short life-time.

\section{Discussion}  

We showed that the B\"ohmer-Harko halos made of an ultralight boson are unstable, that is, we showed that halos fitting galactic rotation curves lose mass very quickly at a rate of half of its mass in a lapse of dozens of millions of years as measured within $r=R$ which is the supposed galactic radius. This behavior is expected since the $\sin r /r$ profile is essentially inconsistent with the ground state attractor solutions. They are different in the first place because the ground state solution in \cite{GuzmanUrena2004,GuzmanUrena2006} are constructed without any assumption of dominance of the self-interaction in the Hamiltonian, in the second place because these solutions satisfy two important conditions, they are bounded  $E<0$ and virialized ($2K+W+3I=0$) and in the third place because the density decays exponentially (not as $\sin r /r$) and is non-zero in the whole space (not cut off at the first zero of the density).

That the total energy of the configurations fitting RCs is positive indicates that the expectation value of the kinetic energy is at least of the same order as the gravitational energy $W$ and the self-interaction energy $I$. This happens in the case of ultralight boson masses used in \cite{RoblesMatos2012}. This is in clear contradiction with the original Thomas-Fermi limit assumed in the original atomic dark matter gas by B\"ohemer and Harko. As explicit cases, using the explicit value in \cite{BoehmerHarko2007} for the boson mass $m=1.44$eV and the parameters obtained in \cite{RoblesMatos2012} for the two galaxies presented here one finds that $\tilde{a}[mt]=2.25\times10^{-10}$ for ESO3050090 and $\tilde{a}[mt]=8.36\times 10^{-11}$ for ESO1870510. Comparing these values with those in our Table, there is a significant difference of seventy orders of magnitude in the scattering length, which may decide between the validity or not of the assumption of dominance of the self-interaction term.



In light of this result, we summarize the status of spherical BEC dark matter halos as follows:

\begin{enumerate}

\item Ground state equilibrium configurations of the GPP system (e.g. \cite{GuzmanUrena2004,GuzmanUrena2006}) are stable in a very general sense, and are late-time attractor solutions, however show insatisfactory RCs. An attempt to solve this problem is for instance the construction of models with mixed ground and excited states \cite{BernalUrena2012}.

\item Stationary configurations constructed using the Madelung tranform in the Thomas-Fermi limit, the B\"ohmer-Harko halos, when considering an ultralight boson, show acceptable RC fits and seem to be a solution to the cusp-core problem in galaxies, however -as we have shown here- they are unstable.

\end{enumerate}

Additional to the instability, the B\"ohmer-Harko halo model (not only when the mass is ultrasmall but in general)  shows an important inconsistency and therefore the applications of the model is in doubt: the radius $R$ is a fitting parameter related to the mass of the boson through the expression $R=\pi \sqrt{\frac{\hbar^2 a}{Gm^3}}$ \cite{BoehmerHarko2007, RoblesMatos2012}. The value of $R$ is a parameter that changes from galaxy to galaxy. This implies that for a given mass of the boson $m$, say the ultralight spinless dark matter model $m \sim 10^{-23}$eV$/c^2$, for a given galaxy there is a fitted $R$ and therefore a scattering length $a$ for that galaxy. However, for the same boson mass, for a different galaxy $R$ is different and therefore $a$ is different. This would be equivalent to consider that bosons interact differently in different galaxies, which is totally unexpected from a dark matter candidate. Conversely, if $a$ is the same for all galaxies, then all the galaxies should have the same $R$ for a fixed $m$ and it is clear that galaxies have different sizes. This is perhaps a problem as important as the problem of instability.

Finally, BEC dark matter shows very interesting properties at cosmic scales, cutting off the mass power spectrum appropriately, and a rather similar behavior to the $\Lambda$CDM model, with the addition that the dark matter candidate is set explicitly to be a boson. This is a strong enough motivation to continue attempting to solve the problem at galactic scale for this dark matter candidate.


\section*{Acknowledgments}

We are thankful to P-H. Chavanis, T. Harko, T. Matos and L. A. Ure\~na-L\'opez for important comments that enriched the paper.
This research is partly supported by grants
CIC-UMSNH-4.9 and CONACyT 106466. 
JJGA and FJRP acknowledge support from CONACyT for qualified graduate programs.


\end{document}